\documentstyle[12pt]{article}
\author{S. Galam and A. Mauger\\ Laboratoire des Milieux
D\'{e}sordonn\'{e}s et H\'{e}t\'{e}rog\`{e}nes\footnotemark[1],\\
Universit\'{e} Paris 6, case 86, 4 place Jussieu, \\75251 Paris
Cedex 05,
   France\\[1ex].} \title{On reducing Terrorism Power: \\A Hint from
Physics}
\date{}

\addtocounter{footnote}{0}
\footnotetext{Laboratoire associ\'{e} au Centre National de la Recherche
Scientifique}
\addtocounter{footnote}{1}

\begin{document}
%%%%%%%%%%%%%%%%%%%%%%%%%%%%%%%%
\def\picture #1 by #2 (#3){
    \vbox to #2{
    \hrule width #1 height 0pt depth 0pt
    \vfill
    \special{picture #3}}}
\def\scaledpicture #1 by #2 (#3 scaled #4){{
\dimen0=#1 \dimen1=#2
\divide\dimen0 by 1000 \multiply\dimen0 by #4
\divide\dimen1 by 1000 \multiply\dimen1 by #4
\picture \dimen0 by \dimen1 (#3 scaled #4)}}
\smallskip
%%%%%%%%%%%%%%%%%%%%%%%%%%%%%%%%%%%%%%
\maketitle
\baselineskip 3.3ex
\footskip 5ex
\parindent 2.5em
\abovedisplayskip 5ex
\belowdisplayskip 5ex
\abovedisplayshortskip 3ex
\belowdisplayshortskip 5ex
\textfloatsep 7ex
\intextsep 7ex

%%%%%%%%%%%%%%%%%%%%%%%%%%%%%
\begin{abstract}
The September 11 attack on the US has revealed an unprecedented terrorism
worldwide range of destruction. Recently, it has been related to the
percolation of worldwide spread passive supporters. This scheme puts the
suppression of the percolation effect as the major strategic issue in the
fight against terrorism. Accordingly the world density of passive
supporters should be reduced below the percolation threshold. In terms of
solid policy, it means to neutralize millions of random passive
supporters, which is contrary to ethics and out of any sound practical
scheme. Given this impossibility we suggest instead a new strategic
scheme to act directly on the value of the terrorism percolation
threshold itself without harming the passive supporters. Accordingly we
identify the space hosting the percolation phenomenon to be a
multi-dimensional virtual social space which extends the ground earth
surface to include the various independent terrorist-fighting goals. The
associated percolating cluster is then found to create long-range ground
connections to terrorism activity. We are thus able to modify the
percolation threshold pc in the virtual space to reach p<pc by decreasing
the social space dimension, leaving the density p unchanged. At once that
would break down the associated world terrorism network to a family of
unconnected finite size clusters. The current world terrorism threat
would thus shrink immediately and spontaneously to a local geographic
problem. There, military action would become limited and efficient.
\end{abstract}

%%%%%%%%%%%%%%%%%%%%%%%%%%%%%
\newpage
Terrorism has existed since centuries. It designates the use of random
violence against civilians in the main purpose to kill them. Until
recently, most terrorist attacks have been performed within restricted
geographic areas connected to the associated terrorism claim. Well-known
European cases of  "local terrorism" are Corsica, Northern Ireland and
Euskadi. In response, nations have developed appropriate intelligence
agencies and special police forces. Various tools have been applied
rather independently by each nation without real multilateral cooperation.

The September 11 attack on the USA has revealed a worldwide range of
action of a given terrorist network.  The response of the United States
to this "long-range terrorism" has been a localized military action,
the success of which is not yet complete. Moreover it is already
understood that follow-up steps are necessary. Nevertheless, defining
these steps is far from being clear. Up to now, a lot of focus was put
on blaming intelligence services for having missed the planning and the
development of the current international terrorism network. As a result,
military and intelligence analysts are looking for the right changes,
which would make these services more efficient. A solution is expected
from an optimization of the conventional tools used so far against local
terrorism.

But at the very moment, there is no available practical scheme to erase the
current worldwide level of terrorism threat. In this work we argue that
the tools used so far to fight local terrorism are inappropriate against
long-range terrorism. We start from a new approach that has been proposed
very recently [1], using the concept of percolation from the Physics of
disorder [2]. Indeed in past years Physics has been proven useful in
shedding new light on a series of social problems [3-6]. According to
this  model, The local and long-range terrorism are the two phases of a
percolation transition in the assembly of randomly distributed passive
supporters to the terrorist cause. Passive supporters designate people
who are not directly involved with terrorism but who would not oppose a
terrorist related act in case they could. Most of them are always dormant.
Some works did analyze the difficulty in mapping covert terrorist
network [7]. The local terrorism corresponds to the disordered phase with
only finite sized clusters of connected people. In that case the density p
of passive supporters is below the percolation threshold pc. The long-range
terrorism corresponds to the case p>pc, the system is ordered with the
existence of an infinite percolating cluster of passive supporters [1].
There, long-range terrorism becomes spontaneously achievable. This range
property is independent of the terrorist net itself.

Within the above frame, the September 11 attack reveals the existence of
a world percolation of passive supporters. Consequently, the reduction of
the world density of passive supporters below the percolation threshold becomes
the major strategic goal of an efficient fight against this international
terrorism. However, even a few percent reduction of the world passive
supporter density would require neutralizing millions of people, either
physically or ideologically, making both options non-ethics and totally
unpractical within reasonable action. Moreover, it has been shown that it
is even not possible to measure the current density of passive supporters
collecting massive and systematic ground information [1].

At this stage, the analysis leads to the very pessimistic conclusion that
there exists no solution to dismiss the current world level of terrorism
threat. The lack of solution comes from the basic idea that, as people are
moving on a two-dimensional surface, this is the space to be considered
while studying their geographic connectivity. This is precisely stated
in the previous percolation model envisioned so far [1], and also implicit
in the approach of the people in charge of the fight against terrorism.
Accordingly, pc is fixed by the ground topology, which is also fixed. Then
the condition p<pc where the long range terrorism is defeated, can be reached
only by changing the density p of the passive supporters. The quandary comes
from the fact that, as stated above, it is both impossible and unacceptable to
neutralize millions of people.

However, in this paper we propose a new strategic scheme to suppress
the passive
supporter percolation without dealing with the passive supporters
themselves. Indeed, a terrorism
cause always produces various independent-fighting goals. We argue
these independent
terrorist-fighting goals can be represented by a set of independent
variables, which in turn
extends the geographic space onto a higher dimensionality virtual
social space. Only passive
supporters populate this space according to their degree of
identification. The associated
percolation occurs there. Once it happens, the virtual percolating
cluster is found to create
additional ground pair links on earth surface, which are found to
extend beyond the original
nearest neighbor distance. The virtual social space dimension
monitors the range of these
additional ground connections. On this basis the percolation
threshold pc in the virtual space
can be modified to reach p<pc by decreasing the social space
dimension, leaving the density p
unchanged.

Already, classical mechanics tell us that the pertinent space to
study the dynamics of a
particle is not the real space in which the elements are actually
moving, but the so-called phase
space in which the particle is reduced to a representative point. The
basic idea is that, at any
given time, the state of the particle is fully determined by the
knowledge of not only its
position, but also its velocity. Then, a figurative point with 6
components, which are 3
coordinates in the real space, and 3 velocity components represents
the particle. Here the
elements we are dealing with are humans and their respective attitude
towards terrorism.
Accordingly, in addition to the geographic coordinates (as considered
in [1]), other social
coordinates, which characterize individual attitude towards the
terrorism under study must be
included. The main terrorism goal, usually the independence of some
geographic area, determines
the first one. Then, as soon as a terrorist group turns active, the
corresponding legal authority
sets on some repressive process against it, setting up a second
social dimension, the
condemnation of the state repression. At this stage, already four
coordinates characterize any
individual in the world. Two determine the position on earth while
the two others measure its
respective concern with respect to the main terrorist cause and the
repression against it. This
set of social parameters may not be complete. Most terrorist groups
relate to additional
dimensions beyond the two mentioned above. For instance, there is
often an ethnic component,
which produces an extra dimension to the representative social space.
The determination of these
parameters will be discussed later on.

To substantiate our proposal, we need to implement our notion of
virtual space, making explicit
the rules to construct it. Considering a discretization of the earth
2-dimensional surface, we
first present a full illustration in the case of a three-dimensional
social space with a
population of 64 persons. For clarity, we choose, as an example, a
simple cubic geometry to
illustrate our purpose, but any other geometry is possible.
Accordingly the 64 persons are
located on an earth square lattice as shown in Figure (1). Among
them, 20 are passive supporters.
Next step is to aggregate the 64 sites by plaquettes of respectively
4 nearest neighbors. It
yields 16 distinct groups.
To keep on a clear visualization we proceed considering only the
upper two 4-sites groups circled
in Figure (1). They are shown in Figure (2). All the 4 sites of a
given plaquette are then
duplicated and distributed along one vertical line, which contains 4
sites as shown in Figure
(3). This vertical direction represents the third dimension, and the
height is an averaged
measure of the degree of identification of the passive supporter to
the terrorist cause. There
exists one vertical site per person. However, only passive supporters
occupy the vertical sites
associated to their respective plaquette, in addition to their ground
sites as shown on Figure
(4). Since many degrees of identification can be met, we postulate
their distribution along the
vertical axis is random. In this process, the image of the 8x8
two-dimensional grid corresponding
to the square on earth in fig. 1 is a 4x4x4 cube in the 3-dimensional
virtual space. The passive
supporters are randomly distributed among the 64 sites of this cubic
lattice. In the virtual
space we restrict interactions to nearest neighbors (nn) as on the
earth lattice. It means that
passive supporters develop an additional connection when both sharing
a similar degree of
identification and belong to nn earth plaquettes. However it is of
importance to notice that nn
connections in the representative space mean earth connections, which
now extend beyond the
original nearest neighbor connections on earth. This is illustrated
in Figs. (4) and (5): The
occupied sites (a)-(f) and (f)-(h) are nearest neighbors in the phase
space, and are then linked
together. However, they are not nearest neighbors on earth so that
these links correspond to
additional bonds on the original two-dimensional plane as shown if Figure (5).

At this stage we have shown explicitly how including a third
dimension creates some additional
connection within the two-dimensional plane, which extends beyond the
original nearest neighbor
connections. Adding other dimensions to the phase space will have the
same effect, i.e. will
create new connections on earth, and increase the range of
interactions on earth between passive
supporters. This is a quite general behavior resulting from the
increase of the number $q$ of
nearest neighbors when the dimension $d$ of the space increases. In
the case of the cubic geometry
chosen to illustrate our purpose $q=2d$. In this case passive
supporters are randomly distributed
on a $d$-dimensional hypercube, which cannot be illustrated on a
figure like in the
three-dimensional case. However, the same construction of the image
of the passive supporters in
the representative space applies, except that the basic unit is
larger than the 64 elements, and
the plaquettes include more than 4 elements in order to keep constant
the hypercubic geometry
when $d$ increases. We can now proceed along presenting the model in
a more general frame.

Assuming roughly a world population of 6.1 billions, we consider a
finite size lattice with $6.1
10^9$ sites and a nearest neighbor connectivity $q_2$. Each lattice
site is occupied by one
person. The earth surface being on a sphere we assume periodic
boundary conditions. We assume a
percentage $p$ of lattice sites occupied by the passive supporters to
the terrorist cause. They
are randomly distributed.

The total surface of the earth is $510$ millions $km^2$, but most of
it covers the oceans, making
the surface of emerged land about $150$ millions $km^2$. The
associated square lattice has then a
side of length $L=12247 km$. For a population of $N=6.1 10^9$ people
on a two-dimensional square
lattice the number of side sites (grid points) is
$N_2=N^{1/2}=78103$. It yields a spacing between
two nearest neighbors of $a_2=L/N_2=157 m$. When going to a
$d$-dimensional space while keeping
the same earth surface, the number of side sites becomes $N_d =
N_1/d$, making the earth spacing
between two nearest neighbors equals to $a_d=L/N_d$. For $d=2, 3, 4,
5, 6 and 7$ we get for the
nearest neighbor spacing in km unit, 0.157, 6.703, 43.823, 135.196,
286.511 and 489.92,
respectively.  The numbers of site sides are 78103, 1827, 280, 91,
43, and 25 respectively.
In simulations, the number $N_7$ used to compute the percolation
threshold is up to 26, quite
close to our finding of 25 [9]. Although the size effects associated
to this finite number of
$N_7$ affect significantly the critical exponents of the percolation
phase transition, they have
little effect on the percolation threshold itself [9]. We also note
that, despite the fact that
the percolation transition on an infinite lattice is of second order,
pre-transitional effects
take place only very close to pc. A variation of p with respect to
pc, although too small to be
detected by intelligence forces, is then sufficient to switch from
the small cluster terrorism to
the worldwide scale of the so-called rich-terrorism. Hence the
difficulty for intelligence
services to predict the event [1].

We are now in a position to discuss our new strategic scheme to
oppose the world wide terrorism
threat without acting on the associated passive supporters. It
articulates on what determines the
value of a percolation threshold, mainly two quantities, the local
connectivity $q$ and the space
dimension $d$. First one measures the number of nearest neighbors for
a given site. We assume it
is a constant between 5 and 20. The second parameter $d$ is the
number of independent variables
needed to localize a point in the associated phase space. It is a
geometric parameter
characteristic of a lattice. Let us now turn to the percolation
threshold. An exact calculation
of pc is available for very few lattices only. Most thresholds are
evaluated from simulations.
However, few years ago we have been able to postulate a quasi-exact
formula, which yields up to
the third decimal all known threshold values [8]. It is a power-law,
which writes for site
dilution,
\begin{equation}
p_c = a[(d-1)(q-1)]^{-b} ,
\end{equation}
with $a=1.2868$ and $b=0.6160$. It indicates that $p_c$ is a
decreasing function of both $d$ and
$q$. This law is universal [8]. We have plotted in Figure (6) the
variation of $p_c$ as a function
of $q$ at dimensions $d=2, d=4$ and $d=10$.

At two dimensions, $p_c$ varies from 0.55 at $q=5$ down to 0.21 at
$q=20$. Therefore, even with
this last very high connectivity of 20, the mean density $p$ of
passive supporters must reach
$21\%$ of the whole population for the associated terrorism to
percolate. It is hard to believe
such a massive support can occur in practice. Most qualitative
estimates mentioned in newspapers
put the level of support to the terrorism cause at around $10\%$. In
that case, at $d=2$ and
$q=20$, it implies that the regions of the phase space where the
local density of representative
points can reach pc are sparse isolated clusters of finite size. But
indeed, at $d=5$ and $q=20$,
we do have $p_c =0.1$. In that case, the terrorist network can thus
percolate with an estimate of
$10\%$ of the population to be passive supporters. Long range
terrorism thus appears possible at
dimension around 5 and not 2. Figure (7) shows the three-dimensional
plane of the percolation
threshold as function of both $q$ and $d$ under two different angles.
It is seen how the threshold
shrinks down to 0.054 at $d=10$ with $q=20$. Already at $d=3$ we have
$p_c=0.36$ at $q=5$ and
$p_c=0.14$ at $q=20$ which is a substantial decrease from
respectively 0.55 and 0.21 but still too
high to be reached.

Above results clearly show that modifying the dimension of the space
has drastic effect on the
value of the percolation threshold. Therefore it allows a strategic
reversal of cracking down the
worldwide terrorist threat, while preserving an ethic attitude, with
an unchanged value of the
density p of passive supporters. And indeed, at contrast with local
terrorism, the novelty of the
current long-range terrorism has been its ability to generate several
additional dimensions to
its representative space. Among others, it embeds a religious
dimension, a social dimension, an
historical dimension and a world bipolarization dimension. It is then
quite realistic to reach
such a high dimension like d=8 giving pc=0.16 and 0.06 for
respectively $q=5$ and 20. Now, only 6
percent of support are enough to get a world percolation cluster, a
small figure that sounds
reasonable. In addition, the nn spacing in the model corresponds to
the maximum range of
interaction on earth (see Fig. 5 for the illustration at $d=3$).
Passive supporters separated by a
distance less than this nn spacing are not necessarily connected
(bond $a-h$ in Fig. 5 for
instance).  However, some of them are indeed connected (bond a-f for
instance) at such a distance
comparable with the size of a state, i. e. few hundreds of
kilometers, for $d\ge 5$. It means
that a few passive supporters can be very efficient at $d\ge 5$, even
if the number of passive
supporters is very low.

Therefore, we can conclude that the relevant strategy to get rid of
massive terrorism is to
decrease $d$, as $p_c$ is very sensitive to this parameter after Eq.
(1) and Figs. (6) and (7).
Formally decreasing q at fixed d would be in principle also efficient
but it would imply in
practice acting on the people, which is not doable. At contrast
decreasing $d$, while out of the
scope of both the physicist investigation and the military action, is
accountable for economical,
cultural and political fields. It does mean that the conventional
tools like intelligence
services have a reduced strategic importance against terrorism. The
basic reason is that, if
politics can successfully reduce the key parameter $d$, they will
only make impossible an infinite
percolating cluster figuring the long-range terrorism, to the benefit
of the situation of finite
isolated clusters for which the conventional tools have proven to be
useful. The war against
terrorism thus requires the simultaneous efforts of both politics to
increase pc and optimized
conventional tools to minimize p within finite size clusters.

In conclusion, we would like to point out that we have listed only
some of the parameters acting
as dimensions in the social space associated to the terrorism
problem. At this stage, the social
space is not yet fully identified. Actually it belongs to Social
Sciences to identify all the
parameters, and define a protocol for their quantitative evaluation.
Still an exhaustive list
will only provide us with a complete set of parameters, which are not
necessarily independent.
The next step will then be to check the independence or quantify
their interaction following the
procedures established in Differential Psychology, this work being
the analog of an
orthogonalization process in mathematics. Only this approach will
allow for the complete
determination of the phase space, through a complete set of
independent variables. The present
paper is then a preliminary work, and we are then eager to see the
community of Social and
Political Sciences to work on this subject within a
multi-disciplinary approach to make it
quantitative. However, let us point out that the model already gives
an important insight on the
winning strategy again the terrorism: attention should not be
focussed on breaking the support
for terrorism. Instead, the winning strategy lies in a reduction in
the dimensions of the social
space in which this support is exerted.

\section{Acknowledgments}
   We would like to thank Dietrich Stauffer for very helpful comment on
making the manuscript clearer.

\newpage
\section{\LARGE References}
1. S. Galam, Eur. Phys. J. B 26, Rapid Note, 269-272 (2002).\\
2. D. Stauffer and A. Aharony, Introduction to Percolation theory,
Taylor and Francis, London
(1994).\\
3. M. BarthÈlÈmy and L. A. N. Amaral, Phys. Rev. Lett. 82, 3180-3183 (1999). \\
4. D. Helbing, I. Farkas and T. Vicsek, Nature 407, 487-490 (2000). \\
5. S. Solomon, G. Weisbuch, L. de Arcangelis, N. Jan and D. Stauffer,
Physica A277 (1-2), 239-247
(2000). \\
6. F. Lilieros, C. R. Edling, L. A. Nunes Amaral, H. E. Stanley and
Y. Aberg, Nature 411,
907-908 (2001). \\
7. V. E. Krebs, Connections 24(3), 43-52 (2002). \\
8. S. Galam and A. Mauger, Phys. Rev. E53, 2177 (1996); Phys. Rev.
E56, 1 (1997). \\
9. D. Stauffer and R. M. Ziff, Int. J. Mod. Phys. C 11, 205 (2000);
A. Ticona and D. Stauffer,
Physica A 290, 1 (2001). \\

%%%%%%%%%%%%%%%%%%%%%%%%%%%%%

\newpage

\section{\LARGE Figure captions}

Fig. 1: A schematic view of 64 persons on a square earth surface with
20 passive
supporters in black circles, others 44 being in gray circle. People
are aggregated into
plaquettes of 4 nearest neighbors as shown for the upper left 16
ones.  \\ [.1 in]
Fig. 2: Nearest neighbor bonds on the earth surface before the
establishment of the
representational social space\\ [.1 in]
Fig. 3: Construction of the third dimension associated to the 8
square lattice sites
of Figure (2). For each plaquette of 4 sites, the associated vertical
line contains 4 sites to
rank the averaged degree of identification to the terrorist goal.\\ [.1 in]
Fig. 4: The representational social space (upper part) with only
passive supporters
occupying their respective sites. Other sites are empty. Below is the
two-dimensional earth
space. \\ [.1 in]
Fig. 5: Earth configuration after the establishment of the
representational social space. Two
more additional links connecting respectively passive supporters (a)-(f)
and (f)-(h) have been
created. Their respective range goes beyond the original nearest
neighbor range.\\ [.1 in]
Fig. 6: Variations of the percolation thresholds as function of the
connectivity q from q=5 to
q=30 at respectively d=2, d=4 and d=10. The horizontal line pc=0.1
shows when percolation occurs
for density p=0.10. \\ [.1 in]
Fig. 7: The three-dimensional plane of the percolation threshold as
function of both q and d
for respectively q=5 to q=20 and d=2 to d=10. The intersection with
the plane at pc=0.10 is
shown. Above the plane, terrorism is localized geographically while
below it is long ranged. \\
[.1 in]

%%%%%%%%%%%%%%%%%%%%%%%%%%%%%%%%%%%%%%%%%%%%%%%%%%%%%%%%%%%%%%%%%%%%%%%
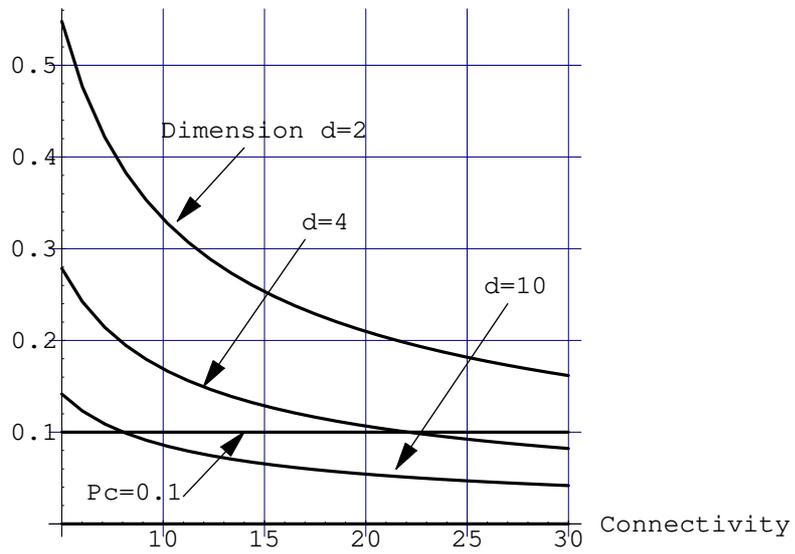
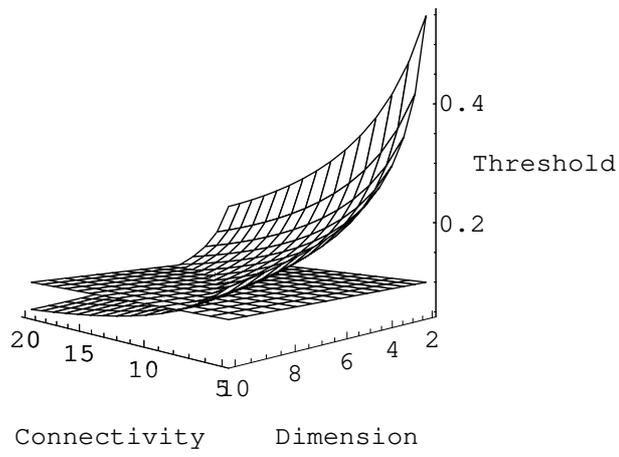
\end{document}